\documentclass[referee]{raa}            

\usepackage{graphicx, times}             
\usepackage{natbib}
\usepackage{amssymb,amsmath}
\bibpunct{(}{)}{;}{a}{}{,}

\usepackage[pagebackref=true]{hyperref}

\usepackage{multirow}

\begin{document}

    \title{Evaluating the spatial intra-pixel sensitivity variations and influence based on space observation 
    }

   \volnopage{Vol.0 (20xx) No.0, 000--000}      
   \setcounter{page}{1}          

    \author{Pei-Pei Wang
        \inst{1,2}
    \and Zi-Huang Cao
        \inst{1,2}
    \and Chao Liu
        \inst{1,2,3,4}
    \and Peng Wei
        \inst{1,2}
    \and Xin Zhang
        \inst{1,2}
    \and Jia-Lu Nie
        \inst{1,2}
    }
   \institute{National Astronomical Observatories, Chinese Academy of Sciences, Beijing 100101, People's Republic of China\\
        \and 
       School of Astronomy and Space Science, University of Chinese Academy of Sciences, Beijing 100049, People's Republic of China\\
       \and
        Institute of Frontiers in Astronomy and Astrophysics, Beijing Normal University, Beijing, 100875, China\\
       \and
        Zhejiang Lab, Hangzhou, Zhejiang, 311121, China\\
    {\it liuchao@nao.cas.cn}\\
   }

\abstract{Intra-pixel sensitivity variations (IPSVs) in charge-coupled devices (CCDs) and complementary metal-oxide-semiconductor (CMOS) detectors constitute a significant source of astrometric error for undersampled stellar observations. 
Since laboratory-based IPSV measurements suffer from limited applicability, we propose a computational method to directly infer IPSV from stellar images and validate it with simulated data.
By minimizing the flux residuals between theoretical and observed stellar models through least-squares fitting, we can successfully recover the IPSV, which is treated as nearly identical across pixels.
Simulations demonstrate that the reconstructed IPSV achieves high accuracy, and the instrumental point spread function (IPSF) restored using this IPSV improves stellar centroiding by nearly 30$\times$, effectively eliminating periodic pixel-phase errors.
The method remains robust under different morphologies of IPSV and varying sampling conditions. Additionally, the framework can be extended to an iterative IPSF–IPSV closed-loop scheme that updates both components simultaneously, providing a practical pathway for continuous detector calibration in future space-based astronomical surveys.
\keywords{instrumentation: detectors; techniques: image processing}
}

\authorrunning{} 
\titlerunning{}  
\maketitle

\section{Introduction}
\label{sect:intro}
Intra-pixel sensitivity variations (IPSV) refer to the non-uniform quantum efficiency distribution within individual detector pixels in charge-coupled devices (CCDs) or complementary metal-oxide-semiconductor (CMOS) imaging arrays. These variations arise primarily from wavelength-dependent photon absorption depths \citep{Refregier_2010} and from spatial inhomogeneities in charge-carrier transport processes such as transmission, diffusion, and scattering within the pixel structure \citep{Kavaldjiev_1997, Kavaldjiev_1998, Kavaldjiev_2001}.
Such effects are intrinsic to the detector design and therefore unavoidable in precision astronomical imaging.

In well-sampled imaging systems, IPSV introduces negligible bias because the stellar point spread function (PSF) is adequately sampled by multiple pixels. In contrast, for space-based observatories where the optical resolution exceeds the detector sampling rate, IPSV produces sub-pixel–dependent errors in PSF sampling, especially when stellar centroids lie away from pixel centers \citep{Skrutskie_1997}. \citet{Anderson_2000} demonstrated using early Hubble Space Telescope (HST) data that IPSV induces pixel-phase errors (PPE) that fundamentally limit astrometric precision. This systematic effect was later mitigated through empirical PSF (ePSF) modeling with dithered observations.
Photometrically, \citet{Kavaldjiev_2001} reported that sub-pixel positioning during dithering can introduce flux variations exceeding 2\%.
Consequently, IPSV represents a major source of systematic error in precision photometry \citep{Mahtani_2015, Cowan_2012}.

Laboratory characterizations of IPSV \citep{Jorden_1994, Kavaldjiev_1997, Kavaldjiev_1998, Kavaldjiev_2001, Piterman_2000, Piterman_2001, Piterman_2002, Toyozumi_2005, Zhan_2017} have been extensively conducted using precision laser-scanning experiments. In these setups, a tightly focused spot is rasterized across target pixels and their neighbors to recover sub-pixel sensitivity maps. 
These measurements reveal that IPSV profiles often follow Gaussian or composite multi-Gaussian morphologies, reflecting combined effects from charge diffusion and pixel architecture.
Notably, IPSV amplitude and morphology show clear wavelength dependence, with back-illuminated CCDs exhibiting smaller variations than front-illuminated designs \citep{Piterman_2000, Piterman_2001, Piterman_2002}. 
More recent efforts have extended such measurements to flight-grade detectors, including Kepler CCD90 \citep{Vorobiev_2019} and various CMOS sensors \citep{Mahato_2017}.

Despite their value, laboratory measurements face intrinsic limitations in reproducing astronomical observing conditions. Key discrepancies arise from finite spot-size, diffraction effects in optical test benches, and environmental turbulence in non-vacuum setups. These factors constrain both the spatial resolution and the fidelity of IPSV measurements, making laboratory results only approximate representations of on-sky performance.
A particularly significant concern is the mismatch in the spectral energy distributions (SEDs) of celestial sources and laboratory illumination, which leads to differing IPSV manifestations. 
These systematic inconsistencies underscore the need for computational approaches capable of inferring IPSV directly from astronomical observations \citep{Zhan+2017}, thereby circumventing the limitations of laboratory characterization.

We present a novel computational method to directly extract IPSV from stellar images. The approach relies on the coupled modulation of the telescope's instrumental PSF (IPSF) and the detector’s IPSV during photon detection. 
The stellar light field is first blurred into a continuous intensity distribution by the IPSF, while the detector samples this distribution discretely, with each pixel’s response further modulated by its IPSV pattern. Consequently, the observed pixel value can be expressed as the integral of the product of the IPSF and IPSV over the pixel area. This relationship provides the theoretical foundation for our IPSV-extraction framework.

The remainder of this paper is organized as follows. 
Section~\ref{sect:IPSV model} introduces the forward modeling approach.
Section~\ref{sect:Simulate observation data} validates the method using simulated observations with known IPSV distributions. 
Section~\ref{sect:results} quantifies photometric and astrometric improvements after IPSV correction, and Section~\ref{sect:discussion} discusses the influence of key model parameters on reconstruction performance.
Conclusions are presented in Section~\ref{sect:conclusion}.

\section{Methods}
\label{sect:IPSV model}

\subsection{Model Foundation}
\label{sect:model}
IPSF characterizes the spatial energy distribution of starlight at the detector focal plane. Ground-based IPSF incorporates both optical characteristics and atmospheric turbulence effects, while space-based IPSF purely reflects the optical response of the system. 
The detector-recorded stellar image results from two successive processes: 
(1) optical blurring described by the IPSF, and 
(2) discrete pixel sampling modulated by the IPSV.
Importantly, IPSV affects the detector's \emph{recorded flux} rather than the 
incident photon count. Although the number of photons arriving at a pixel is 
conserved, different subpixel sensitivities convert the same number of photons into 
different output flux values (in ADU). Consequently, different IPSV patterns lead to 
different total recorded fluxes for a stellar image, even though the incident 
photon flux remains unchanged. This detector-level response nonuniformity enables 
IPSV to be inferred from the discrepancy between the observed and model-predicted 
stellar fluxes.

\begin{figure}[htbp]
\centering
\includegraphics[width=0.8\textwidth, angle=0]{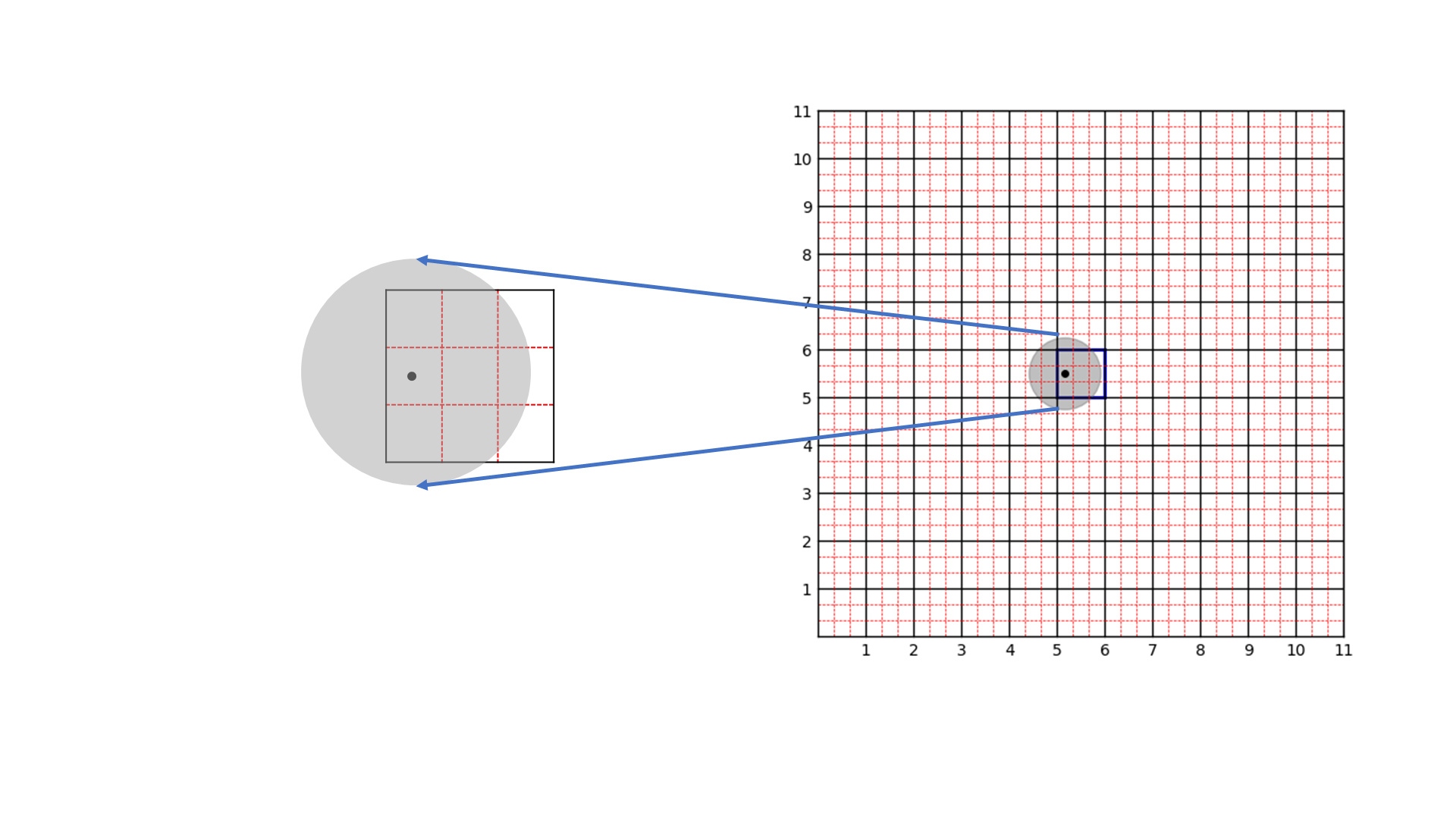}
    \caption{
    Illustration of the pixel and sub-pixel coordinate definitions used in this work. Each detector pixel is indexed by $(i,j)$, and its geometric center is denoted by $(x_{ij}, y_{ij})$. A local Cartesian coordinate system is defined with the origin at the geometric center of the central pixel (row 6, column 6). The continuous coordinates $(x, y)$ denote spatial positions in this local reference frame. 
    Right panel: an $11\times11$ detector pixel array (black lines), where the stellar centroid (black dot) lies within subpixel $(1,2)$ of the central pixel at $(i,j)=(6,6)$. Left panel: a schematic sub-pixel grid (red lines) within a single pixel, showing a $3\times3$ sub-pixel division. 
    }
\label{fig_stardemo}
\end{figure}

Figure~\ref{fig_stardemo} illustrates the geometric configuration adopted in this work. Each stellar image is represented by an $11\times11$ pixel cutout extracted from the detector. All cutouts are recentered such that the stellar centroid lies within the central pixel (row 6, column 6), providing a common reference frame for all stellar images.

For each stellar cutout, we define a local Cartesian coordinate system with origin $O(0,0)$ located at the geometric center of the central pixel. Let $x$ and $y$ denote continuous spatial coordinates in this local reference frame. All spatial quantities in this section are expressed in this coordinate system.
The centroid of the kk-th star is specified by the subpixel offsets $(x_{k}, y_{k})\in[-0.5, 0.5]\times[-0.5, 0.5]$, measured relative to the center of the central pixel.

An arbitrary detector pixel within the stellar cutout is indexed by $(i,j)$, and the coordinates $(x_{ij}, y_{ij})$ denote the geometric center of that pixel in the local coordinate system.
The detector response within each pixel is described by the IPSV, which is defined relative to the pixel center.
A key assumption of this work is that the IPSV pattern is identical for all pixels, such that a single IPSV model applies across the detector.

The flux $f_{ij}^{(k)}$ recorded in pixel $(i,j)$ for the $k$-th star is modeled as the pixel-integrated product of the IPSF and IPSV, plus a term $S_{ij}$,
\begin{equation}
f_{ij}^{(k)}  = f_{*}^{(k)} \iint_{pixel(i,j)} IPSV(x-x_{ij}, y-y_{ij}) \times IPSF(x-x_{k}, y-y_{k})dxdy + S_{ij},
\label{eq1}
\end{equation}
where $f_{*}^{(k)}$ is the intrinsic total flux of the $k$-th star.
Both the IPSV and IPSF are normalized, which satisfy $\iint_{pixel(i, j)}IPSVdxdy=1$ and $\iint IPSFdxdy=1$. 
In practice, the images are assumed to be preprocessed, and only photon shot noise from the stellar signal is included. Accordingly, $S_{ij}$is therefore not explicitly modeled in the noise realization and is retained only for completeness of the forward model. Under these conditions, the impact of this term on the SNR is negligible.

We further subdivide each detector pixel $(i,j)$ into an $m\times m$ regular grid of subpixels. The indices $(u,v)$, with $u,v=1,\dots,m$, specify the subpixel location within a pixel. The quantity $IPSV_{uv}$ denotes the intra-pixel sensitivity at subpixel $(u,v)$ and is assumed to be identical for all pixels and stars. In contrast, the subpixel-integrated IPSF term $IPSF_{uv}^{(k)}$ depends on the stellar centroid $(x_k,y_k)$ and therefore varies from star to star. 

Under this discretization, the flux in pixel $(i,j)$ is computed as the sum of the products of $IPSF_{uv}^{(k)}$ and $IPSV_{uv}$:
\begin{equation}
f_{ij}^{(k)} = f_{*}^{(k)} \sum_{u=1}^{m}\sum_{v=1}^{m}IPSF_{uv}^{(k)}\times IPSV_{uv} + S_{ij}.
\label{eq: single pixel value}
\end{equation}

For the $k$-th star, we denote by $P^{(k)}$ the theoretical stellar flux under a empirical $IPSF^{(k)}$,
\begin{equation}
P^{(k)}(IPSF^{(k)},\, IPSV) = \sum_{i=1}^{n}\sum_{j=1}^{n}f_{ij}^{(k)}.
\label{eq: stellar total flux}
\end{equation}

For real astronomical observations, neither the $IPSF^{(k)}$ nor the $IPSV$ is known a priori. 
Therefore, $T^{(k)}$ denotes the detector-recorded flux of the kk-th stellar 
cutout, extracted directly from the observed image.
Since the detector-recorded flux depends on the underlying $IPSV$, the $IPSV$ can be recovered from $q \,( \ge m^2 )$ stellar images by minimizing the discrepancy between the observed and model-predicted stellar fluxes:
\begin{equation}
\min_{IPSV}
\sum_{k=1}^{q}
\left[
T^{(k)} - P^{(k)}(IPSF^{(k)},\, IPSV)
\right]^2 .
\label{eq:ipvs_lsq}
\end{equation}
Stars with randomly distributed subpixel centroids ensure adequate sampling of 
the $m \times m$ intra-pixel response structure, enabling the IPSV pattern to 
be robustly recovered.

\subsection{Model Assessment Metrics}
\label{sect:Model Assessment}
To comprehensively evaluate the performance of the proposed IPSV extraction 
framework, we adopt a set of quantitative metrics that jointly assess the accuracy of the recovered IPSV and the quality of the reconstructed IPSF.

\begin{enumerate}
\item [(1)] Relative Frobenius Norm (RFN) for IPSV estimation accuracy.

The accuracy of the recovered IPSV pattern, $\hat{\mathrm{IPSV}}$, is quantified using the Relative Frobenius Norm (RFN), a scale-invariant matrix error metric based on the Frobenius norm that measures the normalized discrepancy between $\hat{\mathrm{IPSV}}$ and the ground-truth IPSV \citep{Golub_2013}.
A lower RFN value indicates closer agreement with the ground-truth IPSV. The RFN is defined as
\begin{equation}
RFN = 
\frac{\lVert \hat{IPSV} - IPSV \rVert_F}
{\max\left( \lVert \hat{IPSV} \rVert_F, \lVert IPSV \rVert_F \right)},
\end{equation}
where the Frobenius norm is
\[
\lVert A \rVert_F = 
\sqrt{\sum_{u=1}^{m}\sum_{v=1}^{m} A_{uv}^2 }.
\]
The normalization ensures numerical stability when the IPSV amplitudes are small.
    
\item [(2)] MRE and MAE for reconstructed IPSF.

Given the estimated IPSV $\hat{IPSV}$, the IPSF is reconstructed through inverse modeling to obtain $\hat{IPSF}$. The reconstruction accuracy is evaluated using the mean relative error (MRE) and mean absolute error (MAE):
\begin{equation}
MRE = 
  \frac{1}{N_{\text{pix}}K} 
  \sum_{i,j,k} 
  \frac{p_{i,j,k} - f_{i,j,k}}{f_{i,j,k}},
\label{eq: MRE}
\end{equation}
    
\begin{equation}
MAE = 
  \frac{1}{N_{\text{pix}}K} 
  \sum_{i,j,k} 
  \frac{|p_{i,j,k} - f_{i,j,k}|}{|f_{i,j,k}|},
\label{eq: MAE}
\end{equation}
Here, $f_{i,j,k}$ and $p_{i,j,k}$ denote the ground-truth and reconstructed 
IPSF values at pixel $(i,j)$ for the $k$-th star, $N_{\mathrm{pix}}$ is 
the number of pixels per stellar cutout, and $K$ is the number of stars 
included in the assessment.
    
\item [(3)] Pixel phase error (PPE) analysis.

To further assess the astrometric impact of IPSV, we measure the Pixel Phase Error (PPE), defined as the centroid displacement mainly induced by IPSV. PPE is known to vary sinusoidally with the subpixel phase \citep{Anderson_2000}. Stellar centroids are measured using the windowed centroiding algorithm implemented in SEP \citep{Barbary_2016}, a Python library that reimplements the core functions of Source Extractor \citep{Bertin_1996}.
\end{enumerate}

\section{Data}
\label{sect:Simulate observation data}

\subsection{IPSF simulation}
\label{subsect: Instrument Point Spread Function Simulation IPSF}
For a space-based, two-meter-class optical telescope operating in the absence of atmospheric turbulence, the IPSF is markedly sharper than its ground-based counterpart. Under these conditions, a two-dimensional Gaussian profile provides a practical compromise between analytical simplicity and physical realism, and is therefore adopted instead of the Moffat profile \citep{Trujillo_2001}.

We model the Gaussian widths $\sigma_x$ and $\sigma_y$ as independent 
random variables drawn from $\mathcal{N}(\sigma, (0.05\sigma)^2)$, allowing for up to 5\% relative variation to account for the realistic spatial IPSF variations across the 
detector. 
The IPSF of the $k$-th star is centered at $(x_k,y_k)$ and given by
\begin{equation}
IPSF(x, y) = 
\frac{1}{2\pi\sigma_x\sigma_y}
\exp\left[
-\left(
\frac{(x-x_k)^2}{2\sigma_x^2} +
\frac{(y-y_k)^2}{2\sigma_y^2}
\right)
\right].
\label{ipsf}
\end{equation}
The parameter $\sigma$ is determined from the diffraction-limited 
full width at half maximum via $FWHM = 2.35\sigma$.

\begin{figure}[htbp]
\centering
\includegraphics[width=0.6\textwidth]{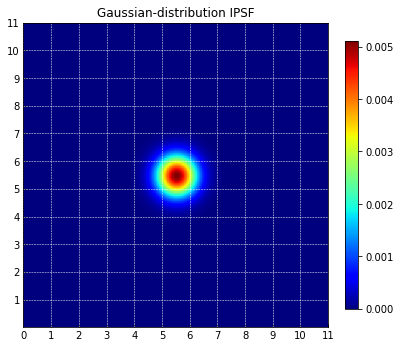}
\caption{Normalized Gaussian IPSF ($\sigma$=0.5 pixel) within 11×11 pixel array. A 1.5-pixel radius from the pixel (6,6) center contains more than 80\% of total flux.}
\label{fig: star PSF.}
\end{figure}

We adopt a unit pixel scale and an $11\times11$ IPSF grid with $\sigma = 0.5$ pixel. Verification confirms that more than 80\% of the flux is enclosed within a 1.5-pixel radius, consistent with the undersampling regime typical for space telescopes. Figure~\ref{fig: star PSF.} shows a normalized example.

In practice, discrepancies exist between observed and theoretical IPSFs due to positional measurement errors. These errors, when coupled with IPSV effects, can propagate to affect the final pixel-integrated values. To quantify this impact, we simulate the measurement error by randomly shifting the above IPSF.
We incorporate normally distributed random measurement errors (obeying $\mathcal{N}(0.02, 0.001)$) along both axes to realistically emulate the positional uncertainty inherent in experimental measurements.

\subsection{IPSV and the stellar objects simulation}
\label{subsect: Intra-pixel Sensitivity Variation Map and Observation Images Simulation}
\begin{figure}[htbp]
\centering
\includegraphics[width=0.8\textwidth]{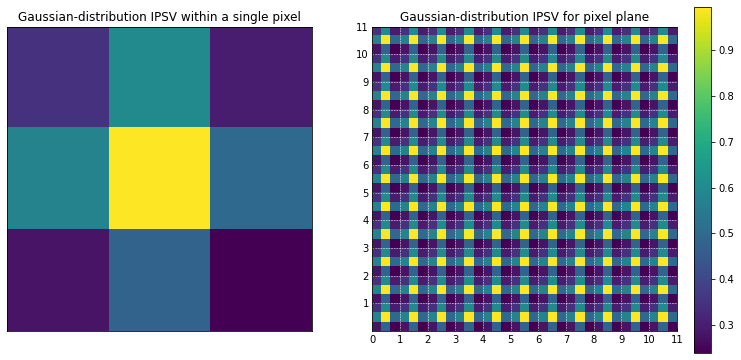}
\caption{
Left: Simulated $3\times 3$ IPSV ($\sigma=0.8$) for a pixel. 
Right: illustration of the resulting IPSVs over an $11\times11$ pixel region when the same IPSV is applied to each pixel.
}
\label{fig:star_IPSV}
\end{figure}

Laboratory measurements indicate that IPSVs typically exhibit Gaussian or composite-Gaussian morphologies. Accordingly, we adopt an asymmetric 
Gaussian IPSV model:
\begin{equation}
IPSV(x, y) = exp\Big(-\frac{(x-\mu_x-i)^2+ (y-\mu_y-j)^2}{2\sigma^2}\Big)
\label{ipsv}
\end{equation}
where $(i,j)$ specifies the pixel center. We set the ground-truth parameters 
to $\mu_x=0.03, \mu_y=0.02$ and $\sigma=0.3$ for the simulation. The left panel of Figure~\ref{fig:star_IPSV} shows the simulated IPSV sampled on a $3\times3$ subpixel 
grid. The right panel presents the corresponding IPSVs, illustrating how the same underlying distribution manifests when aggregated into an $11\times 11$ pixel array. 

As mentioned in Section \ref{sect:model}, stellar image synthesis follows a rigorous physical process, incorporating instrumental effects and noise characteristics. For a 100-second exposure, the observed image is simulated as follows:
\begin{itemize}
    \item [a)] \textbf{Flux scaling.}
    The intrinsic stellar flux is
    \begin{equation}
    flux = 100 \times 10^{(25.83 - mag)/2.5},
    \label{eq_detla_flux}
    \end{equation}
    for a 100~s exposure with zero-point magnitude 25.83.
    
    \item[b)] \textbf{PSF convolution and IPSV modulation.}
    The ideal image is formed by convolving the IPSF with a flux-scaled delta 
    function, followed by multiplicative modulation by the IPSV:
    \begin{equation}
    Image = (IPSF * \delta_{flux}) \;\odot\; IPSV,
    \end{equation}
    where $*$ denotes convolution and $\odot$ denotes element-wise multiplication.

    \item[c)] \textbf{Noise injection.}
    Photon shot noise is injected by applying Poisson sampling to the noise-free model images, with the expected stellar flux in each pixel determined by the input stellar magnitude. Under this assumption, the resulting signal-to-noise ratio is given by $SNR\approx \sqrt{F}$, where $F$ denotes the total stellar flux.
\end{itemize}

We simulate 2,000 stellar cutouts with magnitudes from 18 to 22, corresponding to SNRs of approximately 350 to 60, each with ground-truth IPSF, measured IPSF, and consistent ground-truth IPSV. A randomly selected subset of 1,000 stars is used for IPSV estimation.

\section{Results}
\label{sect:results}
The performance was comprehensively validated through three metrics: (1) IPSV solution accuracy (quantified by RFN), (2) IPSF reconstruction fidelity (assessed via MAE and MRE), and (3) centroiding precision with/without IPSV correction (evaluated using PPE).

\subsection{IPSV solution accuracy and IPSF reconstruction performance}
\label{sect: Methods for evaluating the results}

\begin{figure}[htbp]
\centering
\includegraphics[width=0.8\textwidth]{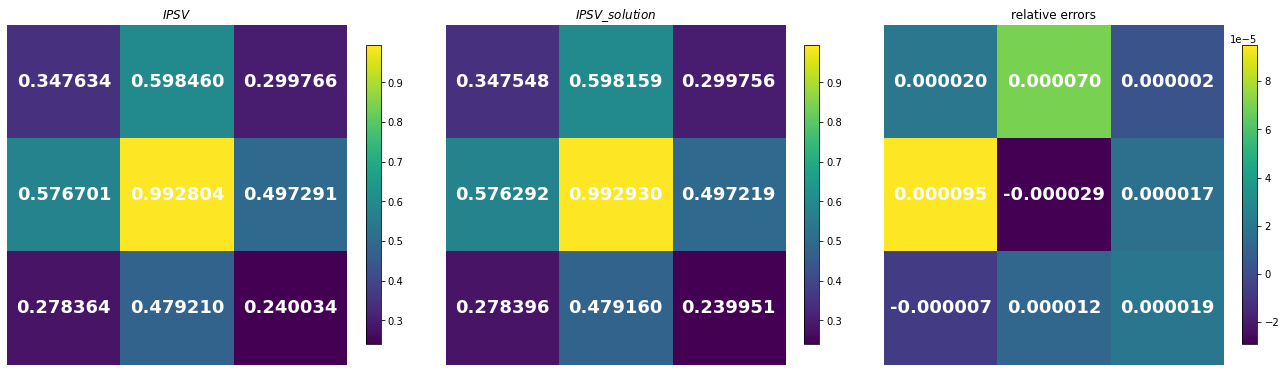}
\caption{
The IPSV solution. 
Left: simulated ground-truth IPSV. 
Middle: IPSV solution $\hat{IPSV}$. 
Right:  relative residual $(IPSV-\hat{IPSV})/IPSV$. 
Numerical annotations in each panel indicate the sub-pixel response efficiency values or error values.}
\label{fig:IPSV.solution.poisson}
\end{figure}

The IPSV solution enables the correction of IPSV-induced distortions in the observed stellar images, thereby allowing the intrinsic IPSF to be accurately reconstructed. 
In this simulation, all removable instrumental noise components (e.g., bias, dark-current residuals, flat-field errors) are excluded so that the IPSV effect can be isolated.
However, Poisson noise is retained, as it represents an irreducible component of real astronomical observations and influences the measured pixel values in conjunction with the IPSV response.
Figure \ref{fig:IPSV.solution.poisson} presents the simulated ground-truth 
IPSV, the IPSV solution $\hat{IPSV}$, and their relative residuals,
\[
R_{uv}=\frac{IPSV_{uv}-\hat{IPSV}_{uv}}{IPSV_{uv}}.
\]
The IPSV solution $\hat{IPSV}$ closely matches the ground-truth IPSV, with maximum relative errors $R_{uv}$ below 0.1\%. The RFN value of $3.44\times10^{-4}$ further confirms the high precision and stability of the IPSV estimation.

\begin{figure}[htbp]
\centering
\includegraphics[width=0.8\textwidth]{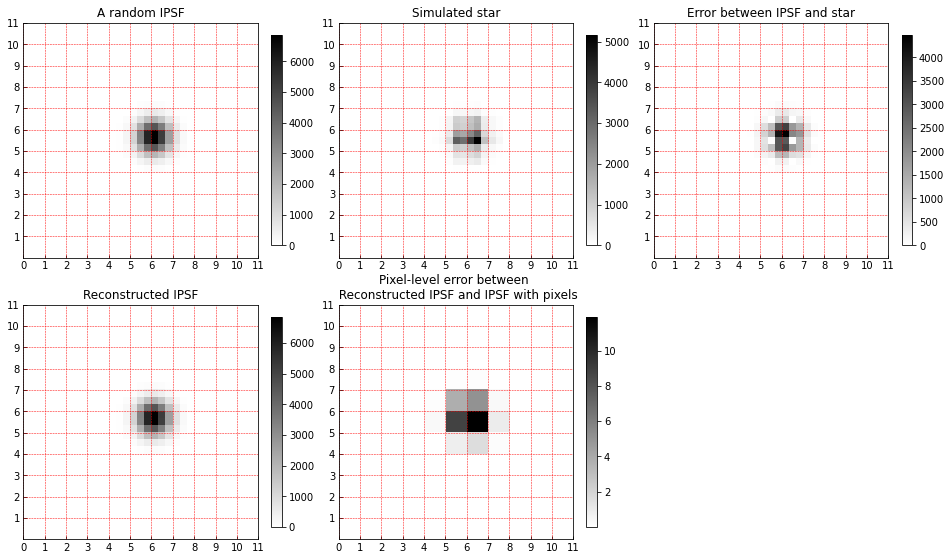}
\caption{
IPSF reconstruction for a representative star.
Top row: 
(1) ground-truth IPSF, 
(2) observed IPSV-modulated stellar image, 
(3) residual between IPSF and stellar image, 
Bottom row:
(4) reconstructed IPSF, 
(5) residual (pixel level) between Reconstructed IPSF and ground-truth IPSF.
}
\label{fig: rePSF one star}
\end{figure}
Using the estimated $\hat{IPSV}$, we subsequently reconstruct the IPSF by removing the IPSV modulation from the stellar image. 
Figure \ref{fig: rePSF one star} illustrates the process for a representative star. 
The top row shows the flux-weighted IPSF, the observed IPSV-modulated stellar image, and their residual, revealing the substantial impact of IPSV on the recorded energy distribution. 
The bottom row presents the reconstructed IPSF together with its residual relative to the ground-truth IPSF, confirming that the IPSV correction successfully restores the lost flux and recovers the IPSF structure.

The reconstruction method demonstrates excellent fidelity, evidenced by an average MRE of $-1.49\times 10^{-4}$ and MAE of $3.1\times 10^{-4}$ across 1000 pairs of reconstructed IPSFs and ground-truth IPSFs. The small MRE value supports negligible systematic bias, while the low MAE indicates tight, consistent agreement with the ground-truth IPSFs, validating both accuracy and robustness of the recovery process.

\subsection{Pixel phase error correction}
\label{sect: The effects of IPSV correction on PPE}
\begin{figure}[h]
\centering
\subfloat[PPE comparison with IPSV $\sigma=0.3$]{
\label{PPE.corrected.noise.IPSF0.3}
\includegraphics[width=0.45\textwidth]{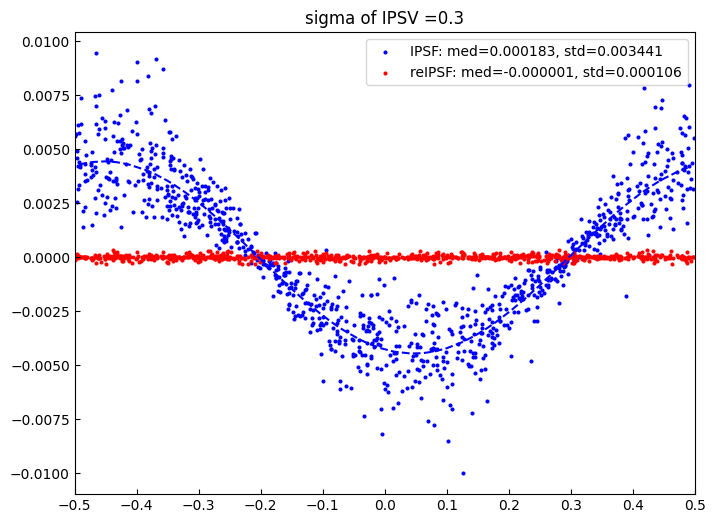}}
\subfloat[PPE comparison with IPSV $\sigma=0.8$]{
\label{PPE.corrected.noise.IPSF0.8}
\includegraphics[width=0.45\textwidth]{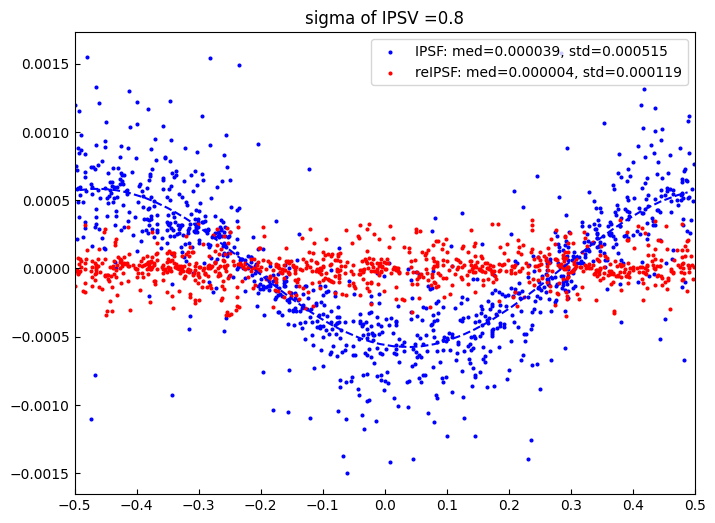}}
\caption{
PPE before and after IPSV correction.  
Blue points and the fitted sinusoidal curve represent PPE measured from the IPSV–distorted stellar images (uncorrected).  
Red points show PPE obtained from the reconstructed IPSF (corrected).  
Left: IPSV with $\sigma=0.3$.  
Right: IPSV with $\sigma=0.8$.
}
\label{fig: PPE corrected results sigma0.3 IPSF0.5}
\end{figure}

Figure \ref{PPE.corrected.noise.IPSF0.3} presents the PPE results for the observed IPSV-modulated stellar image (before IPSV correction) and the IPSF–reconstructed images (after IPSV correction). PPE is plotted as a function of the subpixel phase. 
As expected, the uncorrected measurements exhibit a clear sinusoidal dependence on the pixel phase, a characteristic signature of IPSV-induced centroiding errors.

For the case of $\sigma=0.3$, IPSV correction suppresses the PPE amplitude from 
approximately $5\times10^{-3}$ pixel to below $1.0\times10^{-5}$ pixel.
Quantitatively, the median centroiding error decreases from $1.8\times10^{-4}$ to 
$1.0\times10^{-6}$, while the standard deviation is reduced from $3.4\times10^{-3}$ to $1.1\times10^{-4}$, representing a factor of $\sim30$ improvement.
The sinusoidal pixel-phase modulation is effectively removed, demonstrating that the IPSV correction substantially restores centroiding linearity across the pixel.

A similar level of improvement is obtained for the broader IPSV distribution 
($\sigma=0.8$), as shown in Figure~\ref{PPE.corrected.noise.IPSF0.8}.
The residual PPE is slightly larger than in the $\sigma=0.3$ case, which is expected because a wider IPSV profile averages over finer subpixel variations, reducing high-frequency structure and slightly weakening the phase-dependent signature.
Nevertheless, the corrected PPE remains more than an order of magnitude smaller than the uncorrected measurements.

\section{Discussion}
\label{sect:discussion}
The experimental validation of IPSV correction method, based on Gaussian PSF and asymmetric Gaussian IPSV, demonstrates significant improvements in photometric and astrometric precision. By discretizing the detector area into $11\times 11$ pixel grid with each pixel subdivided into $3\times 3$ subpixels, we rigorously quantified the impact of IPSV on flux loss and centroiding accuracy. The results show that the proposed IPSV correction simultaneously mitigates flux depletion and enhances centroid measurement precision.

Our experimental framework incorporates flexible parameter designs to systematically evaluate the algorithm's performance under varying conditions. Specifically, we will examine: (1) different subpixel configurations, (2) varying numbers of stellar samples in the computation, and (3) diverse IPSV morphological profiles. Throughout these tests, we maintained the ground-truth IPSV with $\sigma=0.3$ and the original Gaussian PSF profile to isolate parameter effects under typical observational conditions.
In addition, a broader IPSV distribution with $\sigma = 0.8$ is tested in Section \ref{sect: Subpixel resolution effects} and found to exhibit the same qualitative behavior as the $\sigma = 0.3$ case; subsequent analyses therefore focus on $\sigma = 0.3$.

\subsection{Subpixel resolution effects}
\label{sect: Subpixel resolution effects}
Theoretically, increasing the subpixel sampling resolution allows the IPSV distribution to be represented with greater spatial fidelity and enables the model to resolve finer intra-pixel variations. However, a higher subpixel resolution also introduces a larger number of free parameters, increasing both computational cost and the sensitivity of the inverse problem to noise. To quantify this trade-off, we solved for the IPSV and reconstructed the IPSF using subpixel grids ranging from $3\times 3$ to $15\times 15$ per pixel.

\begin{figure}[htbp]
  \centering
  \includegraphics[width=0.8\textwidth]{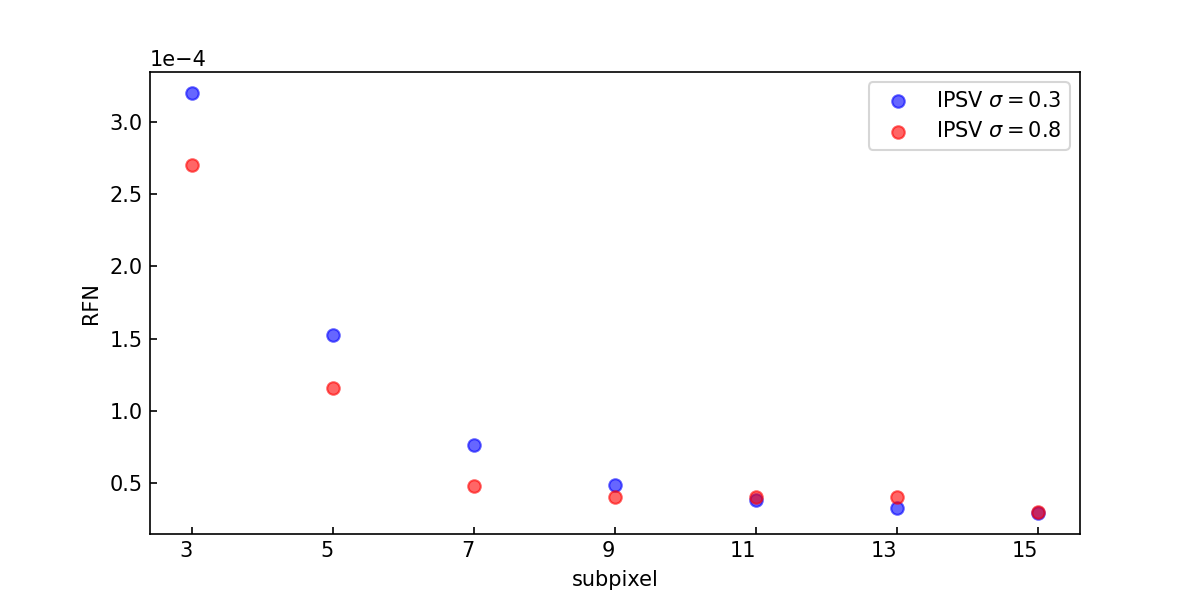}
\caption{$\hat{IPSV}$ accuracy (RFN value) vs. subpixel resolution.}
\label{MREMAE.IPSV.SUBPIXELS_0308}
\end{figure}

Figure~\ref{MREMAE.IPSV.SUBPIXELS_0308} shows that the IPSV accuracy improves systematically with increasing subpixel resolution for both $\sigma = 0.3$ and $\sigma = 0.8$.
Although RFN values differ slightly between the two cases, their overall trends are qualitatively identical. Therefore, for clarity and without loss of generality, the subsequent analysis focuses on the representative case of $\sigma = 0.3$.

For $\sigma=0.3$, the RFN decreases from $3\times10^{-4}$ at $3\times3$ resolution to below $5\times10^{-5}$ at $15\times15$, representing a nearly sixfold enhancement. The MRE and MAE values in Table~\ref{table: MRE/MAE} exhibit consistent decreasing trends, with MRE decreasing from $2.19\times10^{-4}$ to $2.7\times10^{-5}$ and MAE from $3.61\times10^{-4}$ to $4.6\times10^{-5}$. 

\begin{table}[h]
\caption{MRE/MAE with various subpixels.}
\centering
\begin{tabular}{|c|c|c|c|c|c|c|c|}
\hline
& $3\times 3$ & $5\times 5$ & $7\times 7$ &$9\times 9$ & $11\times 11$ &$13\times 13$ &$15\times 15$ \\
\hline
MRE & $2.19\times10^{-4}$ &  $1.20\times10^{-4}$ 
    & $6.6\times10^{-5}$ & $4.4\times10^{-5}$ 
    & $3.7\times10^{-5}$ & $3.2\times10^{-5}$ 
    & $2.7\times 10^{-5}$\\
MAE & $3.61\times10^{-4}$ &  $1.69\times10^{-4}$ 
    & $8.3\times10^{-5}$ & $7.1\times 10^{-5}$ 
    & $5.8\times 10^{-5}$ & $6.2\times 10^{-5}$ 
    & $4.6\times 10^{-5}$\\
\hline
\end{tabular}
\label{table: MRE/MAE}
\end{table}

The rate of improvement, however, saturates beyond $9\times9$. Between $9\times9$ and $15\times15$, both MRE and MAE decrease only marginally, and MAE even shows slight fluctuations due to noise amplification. This behavior suggests that the inverse problem becomes progressively ill-conditioned at higher resolutions, where the number of IPSV parameters ($m^2$ per pixel) grows faster than the number of independent stellar constraints. 
Consequently, noise sources such as photon shot noise, readout noise, flat-field uncertainties, and interpolation artifacts become progressively more influential, limiting the practical benefits of extremely high subpixel resolutions.
PPE correction exhibits a similar dependence on subpixel resolution. The standard deviation of PPE decreases from $3.4\times10^{-3}$ pixel at $3\times 3$ resolution to below $10^{-6}$ pixel at $9\times 9$, after which further improvements become negligible.

\begin{figure}[htbp]
  \centering
  \includegraphics[width=0.8\textwidth]{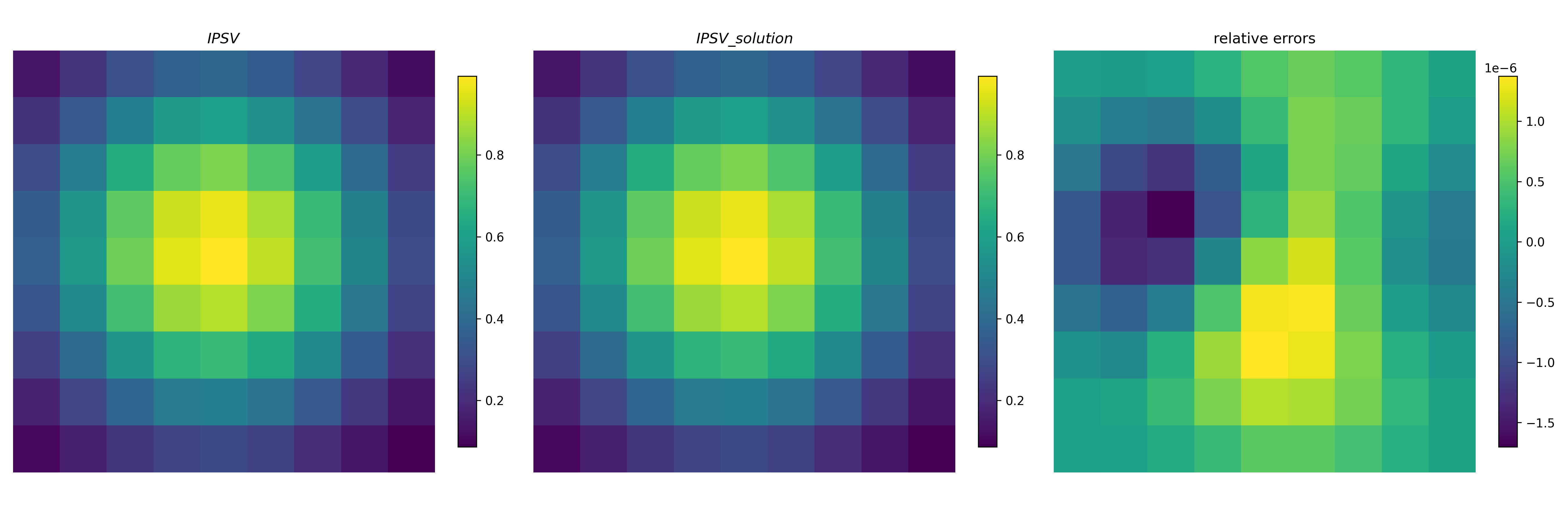}
\caption{
IPSV solution at the optimal $9\times9$ subpixel resolution.  
Left: simulated ground-truth IPSV. 
Middle: recovered IPSV $\hat{IPSV}$. 
Right: corresponding relative residual.
}
\label{IPSV.SUBPIXELS9}
\end{figure}

Figure~\ref{IPSV.SUBPIXELS9} illustrates the IPSV reconstruction at the optimal $9\times9$ subpixel resolution, showing close agreement between the recovered and ground-truth IPSV distributions. At this resolution, sufficient stars are available to constrain the IPSV model (with $m^2=81$ parameters), while avoiding noise-driven instability.

\subsection{Numerical Stability of IPSV Solutions vs. Stellar Count}
\label{sect: The effect of the number of simulated stars}
The precision of IPSV solutions depends on the number of stellar images used in model fitting. To quantify this dependence, we incrementally increased the sample size from 10 to 1000 stellar images (in increments of 10) and analyzed the resulting variations in accuracy. 
For each sample size, the experiment was repeated 100 times with random draws from the full 2000-star dataset, and the mean RFN across all repetitions was adopted as the performance indicator.

\begin{figure}[htbp]
  \centering
  \includegraphics[width=0.8\textwidth]{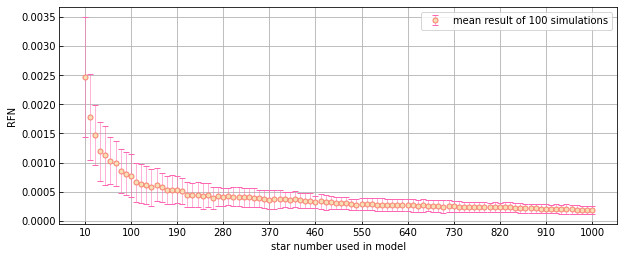}
  \caption{IPSV solution convergence with stellar counts. Error bars show 2$\sigma$ variation across 100 trials.}
  \label{Trend of RFN for IPSV final solution}
\end{figure}

The analysis reveals a nonlinear relationship between the number of reference stars and the IPSV solution. The results demonstrate that utilizing 400 reference stellar images achieves 90\% of the maximum attainable accuracy (RFN $\approx 5\times10^{-4}$), while increasing the sample size to 1,000 stellar yields only a smaller 
incremental improvement (RFN $\approx 3.5\times10^{-4}$), as illustrated in Figure \ref{Trend of RFN for IPSV final solution}. 
This diminishing-return behavior indicates an optimal operating regime in which additional stars contribute progressively less to the overall solution precision.

\subsection{IPSV morphology robustness}
\label{sec: morphology}
\begin{figure}[ht]
  \centering
  \subfloat[Three-Gaussian IPSV surface.]{
\label{IPSV.TG.SURF}
\includegraphics[width=0.45\textwidth]{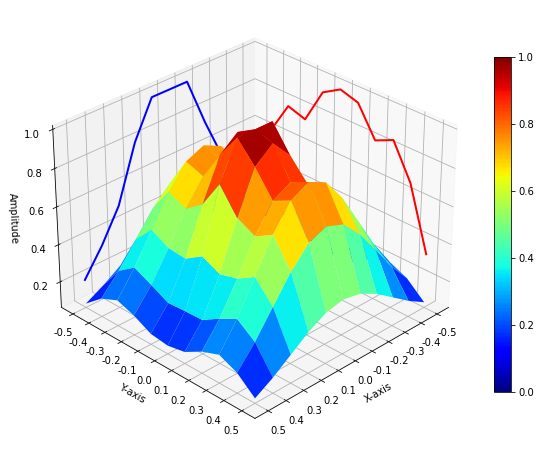}}
\subfloat[Three-Gaussian IPSV map.]{
\label{IPSV.TG.MAP}
\includegraphics[width=0.45\textwidth]{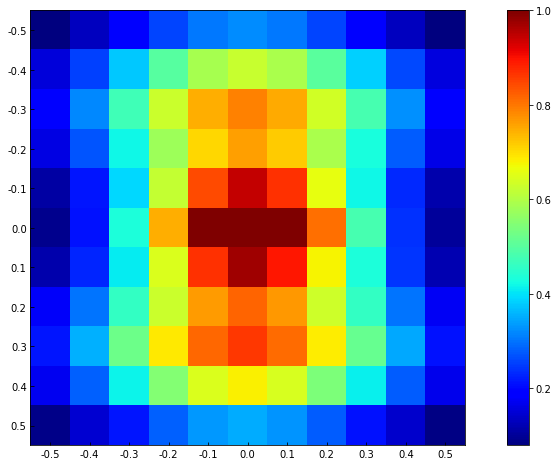}}
\caption{The plane projection of the three-Gaussian IPSV in the z-axis direction (left) and the pixel array (right).}
\label{TG_IPSV_map_3D_surf}
\end{figure}

Our experimental framework initially employs a Gaussian IPSV model as the baseline configuration. Nevertheless, pixel-response measurements based on pixel-array scanning have demonstrated that some devices exhibit significantly more complex intra-pixel structures.
In particular, \citet{Toyozumi_2005} reported IPSV profiles featuring a primary peak flanked by two weaker lateral lobes, well-characterized by a three-Gaussian decomposition. To evaluate the robustness of our method, we simulate an IPSV configuration following this three-Gaussian morphology (Figure~\ref{TG_IPSV_map_3D_surf}).

More complex IPSV shapes substantially amplify the uncorrected PPE: in our simulations, the raw PPE increases by a factor of 5–10 relative to the single-Gaussian case owing to the stronger modulation of the flux distribution within the central pixel. Despite this increase in IPSV, our correction remains highly robust—after calibration, the residual centroid errors are consistently suppressed to below $10^{-4}$ pixels across this morphology. This demonstrates that although IPSV morphology significantly affects the uncorrected response, the correction process can recover high-precision astrometry provided that IPSV is explicitly modeled and solved.

\subsection{IPSF–IPSV Alternating Reconstruction Framework for Real Observations}
Applying IPSV inversion directly to real astronomical images requires addressing the fundamental coupling between the optical IPSF and the intra-pixel sensitivity variation. Because the detector records only pixel-integrated flux, IPSF and IPSV cannot be independently recovered from a single observation; instead, they must be constrained jointly. We therefore propose an alternating reconstruction framework designed to iteratively decouple and recover both quantities.
Here, we do not provide a proof of convergence for the alternative solution; instead, this section focuses on demonstrating feasibility rather than on performance optimization.

In practical applications, the initial value of IPSV is measured in the laboratory, while the initial IPSF is obtained using the PSFEx or ePSF method.
Using this updated IPSV, the observed stellar images can be demodulated to extract a more accurate IPSF, which serves as the input to the next iteration. Repeating the sequence of IPSV inversion and IPSF reconstruction forms a closed-loop update scheme,
\begin{equation}
\begin{aligned}
&\text{IPSF}_{\text{n+1}} = f(\text{IPSV}_{\text{n}}, \text{Observations}), 
&\text{IPSV}_{\text{n+1}} = g(\text{IPSF}_{\text{n}}, \text{Observations}),
\end{aligned}
\end{equation}
which progressively reduces the IPSF–IPSV degeneracy and drives the system toward a mutually consistent solution. With sufficient sampling of pixel phases and adequate SNR, this alternating optimization enables simultaneous recovery of the IPSF and the IPSV morphology. The framework thus provides a practical and physically grounded pathway for extending our IPSV reconstruction method from controlled simulations to real observational data.

\section{Conclusion}
\label{sect:conclusion}

We have presented a physically motivated and computationally efficient method for extracting IPSV directly from stellar images. By modeling stellar image formation as the pixel-integrated product of the IPSF and the IPSV response, our approach leverages randomly sampled subpixel positions to recover the underlying IPSV from observational data.

Using simulated stellar images based on Gaussian IPSFs and asymmetric Gaussian IPSVs, we demonstrated that the IPSV distribution can be accurately reconstructed through a least-squares inversion. The recovered IPSV exhibits relative errors below 0.1\% and an RFN of $3.4\times10^{-4}$ with respect to the ground-truth. The corresponding IPSF reconstruction successfully restores the IPSV-induced flux deficits and improves centroiding accuracy by more than an order of magnitude, suppressing pixel phase errors from $\sim 3.4\times10^{-3}$ to $\sim 10^{-4}$ pixels.

A systematic evaluation across subpixel resolutions, stellar sample sizes, and IPSV morphologies shows that IPSV inversion remains numerically stable and highly accurate under a broad range of configurations. Optimal performance is reached at $9\times9$ subpixel sampling, with diminishing returns at higher resolutions, while approximately 400 stellar samples provide sufficient constraints for high-fidelity recovery. The method also remains robust for complex multi-Gaussian IPSV profiles, consistently achieving sub-$10^{-4}$ centroiding precision.

Finally, we outline an alternating IPSF–IPSV reconstruction framework that 
iteratively updates both quantities, offering a practical pathway for applying this technique to real astronomical observations. This work demonstrates the 
promise of IPSV inversion as a calibration tool for precision photometry and 
astrometry in future space-based surveys.

\begin{acknowledgements}
The authors acknowledge support by the China Manned Space Project (Nos. CMS-CSST-2025-A19, CMS-CSST-2025-A21), and the National Natural Science Foundation of China (NSFC) (No.12073047). The sincere thanks go to Dr.Hugh R. A. Jones (Center for Astrophysics, University of Hertfordshire, Hatfield, UK) for his valuable suggestions and revisions to the paper.
\end{acknowledgements}

\bibliographystyle{raa}
\bibliography{bibtex}

\end{document}